\documentclass[12pt]{iopart}
\usepackage{iopams} 
\usepackage{dsfont}
\usepackage{braket}
\usepackage{tikz}
\newcommand{\tikzcircle}[2][red,fill=red]{\tikz[baseline=-0.5ex]\draw[#1,radius=#2] (0,0) circle ;}%

\usepackage[colorlinks=true,linkcolor=blue,urlcolor=blue,citecolor=blue,pdfusetitle]{hyperref}

\usepackage{cite}
\expandafter\let\csname equation*\endcsname\relax

\expandafter\let\csname endequation*\endcsname\relax
\usepackage{amsmath}

\begin{document}

\title[Dissipative quantum many-body dynamics in (1+1)D QCA and QNNs]{Dissipative quantum many-body dynamics in (1+1)D quantum cellular automata and quantum neural networks}
\author{Mario Boneberg$^1$, Federico Carollo$^1$ and Igor Lesanovsky$^{1,2}$}
\address{$^1$ Institut f\"ur Theoretische Physik, Universit\"at Tübingen, Auf der Morgenstelle 14, 72076 T\"ubingen, Germany}
\address{$^2$ School of Physics and Astronomy and Centre for the Mathematics and Theoretical Physics of Quantum Non-Equilibrium Systems, The University of Nottingham, Nottingham, NG7 2RD, United Kingdom}

\begin{abstract}
Classical artificial neural networks, built from perceptrons as their elementary units, possess enormous expressive power. Here we investigate a quantum neural network architecture, which follows a similar paradigm. It is structurally equivalent to so-called (1+1)D quantum cellular automata, which are two-dimensional quantum lattice systems on which dynamics takes place in discrete time. Information transfer between consecutive time slices --- or adjacent network layers --- is governed by local quantum gates, which can be regarded as the quantum counterpart of the classical perceptrons. Along the time-direction an effective dissipative evolution emerges on the level of the reduced state, and the nature of this dynamics is dictated by the structure of the elementary gates. We show how to construct the local unitary gates to yield a desired many-body dynamics, which in certain parameter regimes is governed by a Lindblad master equation. We study this for small system sizes through numerical simulations and demonstrate how collective effects within the quantum cellular automaton can be controlled parametrically. Our study constitutes a step towards the utilisation of large-scale emergent phenomena in large quantum neural networks for machine learning purposes.
\end{abstract}

\maketitle

\section{Introduction}
\begin{figure}[t]
    \centering
    \includegraphics[width=\textwidth]{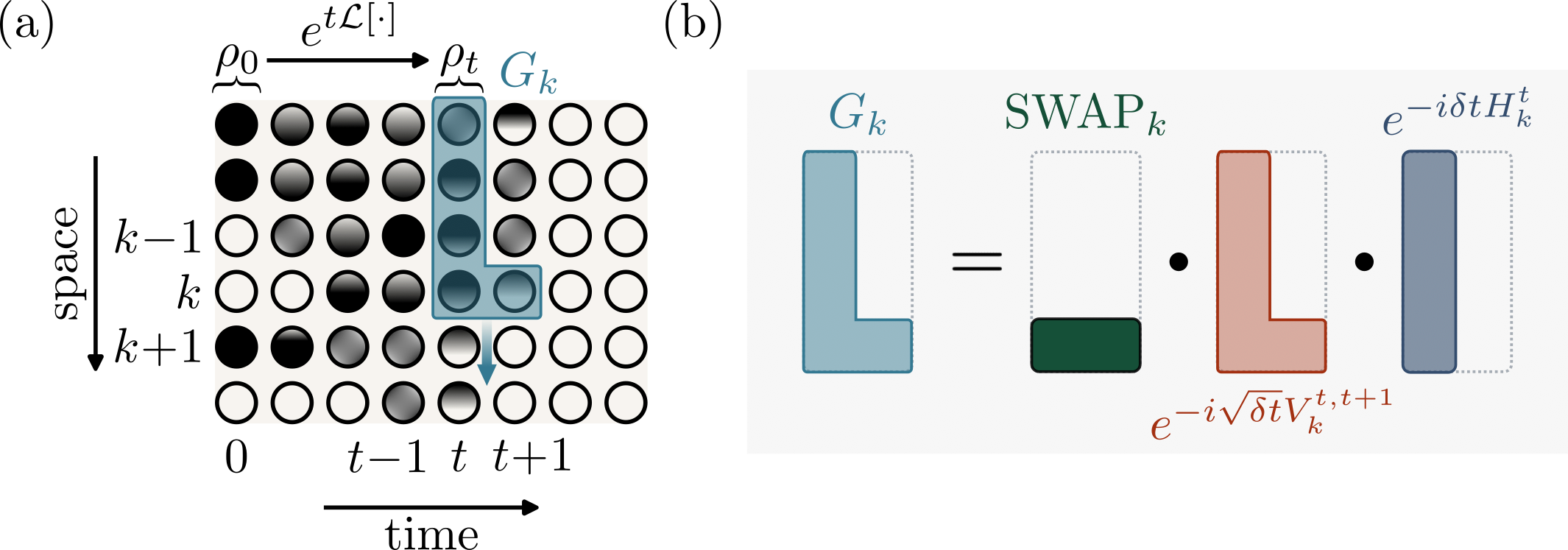}
    \caption{\textbf{(1+1)D Quantum Cellular Automaton architecture.} (a) Two-level systems which can be found in the occupied (circle with black filling) or vacant (circle with white filling) state or superpositions of them are organized into a two-dimensional lattice. The vertical axis can be interpreted as indicating a spatial dimension and consists of $N$ sites of which we show only a subset of six here. With respect to the horizontal axis we label the first layer by $0$. This layer contains the initial configuration $\rho_0$ and all other sites are in the vacant state. By means of the gate $G_k$, sequentially applied along the spatial direction, the initial state is then propagated to $t$ by successively updating two adjacent layers. The horizontal axis can thus be effectively seen as a time-axis. In the main text we show how suitably chosen gates $G_k$ can give rise to a generic Lindbladian evolution of the initial state $\rho_t = e^{t \mathcal{L}}[\rho_0]$. Note that the $G_k$ may act nontrivially on the current layer such that, for instance, the state of layer $0$ after their application is in general different from $\rho_0$. However, for sake of illustration, we neglect such subtleties. (b) The gate $G_k$ can be written as a product of simpler gates as depicted. Our analysis goes beyond the case where $G_k$ acts on only four \textit{control sites} in layer $t$ and applies to general local Hamiltonians and dissipative processes.}
    \label{qca}
\end{figure}

Developing powerful and at the same time tailored computational models is of great relevance in both quantum simulation and quantum machine learning \cite{georgescu2014,altman2021,schuld2022, schuld2014, schuld2021,biamonte2017,wilkinson2022, poland2020, wolpert1997}. One particular computational paradigm is provided by quantum cellular automata (QCA) \cite{arrighi2019, farrelly2020}. As a quantum generalization of classical cellular automata \cite{vonneumann1996, wolfram1983} they are usually characterized by a discrete-time, local evolution of an ensemble of identical finite-dimensional quantum systems via a translationally invariant unitary operator, although many different versions exist \cite{piroli2020,duranthon2021,hillberry2021,mlodinow2021,nigmatullin2021,piroli2021,ney2022,sellapillay2022,shirley2022}. Experimental progress in controlling atomic lattice systems with the ability of single-atom addressing \cite{labuhn2016, bernien2017, kim2018,browaeys2020, wintermantel2020, ebadi2021, scholl2021} has stimulated research into so-called (1+1)D quantum cellular automata \cite{lesanovsky2019,gillman2020,gillman2021a,gillman2021b,gillman2022a,gillman2022b} (see Fig.~\ref{qca}). Being organized into layers of finite-dimensional quantum systems, with a dynamics implemented by the sequential application of unitaries supported on two adjacent layers, they realize one spatial dimension and one effective time dimension. Originally introduced as quantum versions of the classical Domany-Kinzel cellular automaton \cite{domany1984, lesanovsky2019}, these models in the past have given rise to studies regarding the impact of quantum effects on universal behaviour in out-of-equilibrium critical dynamics \cite{lesanovsky2019,gillman2020,gillman2021a,gillman2021b,gillman2022a,gillman2022b}. \\
Dynamics can be often associated with the processing of data. This is the case, for instance, for Hopfield neural networks \cite{hopfield1982,amit1985a,amit1985b,amit1987}, in which given classical spin configurations can be retrieved by performing a dynamics which minimizes a suitable energy function starting from an initially presented configuration. Another example in which patterns can be retrieved starting from data contained in an initial state is given by modern feed-forward neural networks \cite{nielsen2015,goodfellow2016}. Here layers of computational units --- the perceptrons --- propagate low-level representations of the data according to their characteristic properties in order to output higher-level representations. \\
More recently, much effort has been invested in introducing quantum effects into pattern recognition tasks \cite{killoran2019,mangini2021,bravo2022,fiorelli2022,rotondo2018}. Exploiting coherence and entanglement it is believed that such approaches may recognize 'atypical' patterns and could pave the way towards a quantum advantage in machine learning \cite{biamonte2017}. Of particular relevance for the investigation in this paper are so-called dissipative quantum neural networks (QNNs) \cite{beer2020,sharma2022}. Having highly modular structures, these are instances of quantum deep learning architectures and---analogously to classical deep neural networks with layers of neurons \cite{lecun2015,nielsen2015,goodfellow2016}---are composed of multiple layers of qudits. A perceptron within this architecture is realized as a unitary operator which acts on the qudits of two adjacent layers.\\
Being both quantum lattice models with successive unitary layer-to-layer dynamics already suggests that (1+1)D QCA and dissipative QNNs are similar concepts and recently their equivalence was established \cite{gillman2022b}. Moreover, it was also shown that, by appropriately choosing the unitary gates, the open quantum dynamics of nonequilibrium critical models   \cite{carollo2019} may be approximated \cite{gillman2022b}. Regardless of the unitary nature of QCA, a dissipative evolution may indeed emerge at the level of the reduced states of single layers. Indeed, since they naturally implement a many-body version of so-called collision models \cite{lorenzo2017,ciccarello2017,ciccarello2022, cattaneo2021, cattaneo2022}, (1+1)D QCA and QNNs are closely related to Markovian open quantum dynamics.\\
In this work we provide a systematic study of how (1+1)D QCA dynamical rules can give rise to emerging Lindbladian dynamics. More precisely, we show, in the limit of large number of layers, how to decompose  the layer-to-layer unitaries into suitable  local gates in such a way that a desired Lindblad dynamics can be approximated \cite{lindblad1976, gorini1976, breuer2002}. We find such a decomposition for arbitrary local Hamiltonians and jump operators, allowing for a full parametric control over local coherent and dissipative contributions. As outlined above, our results are relevant to the controlled investigation of quantum many-body systems, but may also offer a route towards  quantum-enhanced versions of machine learning tasks. We illustrate our results by implementing two different versions of open quantum Ising models \cite{rose2016,overbeck2017} and an open quantum reaction-diffusion model \cite{vanhorssen2015} on the (1+1)D QCA platform. At large scales, these models are known to feature emergent collective behavior, a necessary requirement for, e.~g., ergodicity-breaking  phenomena which enable to establish a pattern retrieval dynamics analogous to the one of Hopfield neural networks.
\section{(1+1)D Quantum Cellular Automaton and local update rules}
We define our QCA on a two-dimensional lattice with $N$ vertical and $L$ horizontal sites \cite{lesanovsky2019,gillman2020,gillman2021a,gillman2021b,gillman2022a,gillman2022b} (see Fig.~\ref{qca}). With each lattice site we associate a two-level quantum system with Hilbert space generated by the basis states $\ket{\tikzcircle[black, fill=white]{2pt}}$ and $\ket{\tikzcircle[black, fill=black]{2pt}}$, referred to as \textit{vacant} and \textit{occupied} state, respectively. The Hilbert space of the total system is constructed as the tensor product of the two-level Hilbert spaces. At time $t=0$, the first column is chosen to be in some initial configuration with the other sites being in the vacant state. We denote the corresponding initial state of the total system as $\ket{\Psi_0}$. A dynamics can then be introduced on the full lattice by successively applying global unitary update-operators which only act nontrivially on two adjacent columns, starting from the first column and proceeding towards the right, as shown in Fig.~\ref{qca}(a). Since each of these consecutive update steps causes a nontrivial action on one new column to the right, the horizontal axis of the lattice becomes an effective time axis. Thus, labeling the columns as $0, 1, ...,t,t+1,...,L-1,L$, we may denote the global update-operators as $\mathcal{G}_{t,t+1}$ and assume that they are the same for all $t$ (we will later study the effect of relaxing this assumption). The state of this (1+1)D QCA (one spatial dimension and one effective time dimension) at time $t+1$ is then
\begin{equation*}
    \ket{\Psi_{t+1}} = \mathcal{G}_{t,t+1} \cdots \mathcal{G}_{0,1} \ket{\Psi_0}.
\end{equation*}
Equivalently, the dynamics of this model may be expressed through a recurrence relation on the level of the reduced states at two consecutive times. The latter are obtained by tracing out, in the instantaneous density matrix, the part of the system not pertaining to the corresponding time-column. Indeed, denoting by $\Tr_{\neq t+1}(\cdot)$ and $\Tr_{< t+1}(\cdot)$ the trace over all sites apart from those in column $t+1$ and the trace over the sites in columns $0,1,...,t$, respectively, we see that
\begin{align}\label{eq1}
        \rho_{t+1} &= \Tr_{\neq t+1}(\ket{\Psi_{t+1}}\bra{\Psi_{t+1}})  \nonumber \\
            &= \Tr_{< t+1} (\mathcal{G}_{t, t+1} \big( \ket{\Psi_{t }}\bra{\Psi_{t}}\otimes \ket{0}_{t+1} \bra{0} \big) \mathcal{G}_{t, t+1}^\dagger)  \nonumber \\
            &= \Tr_{t} (\mathcal{G}_{t, t+1} \big( \Tr_{<t } (\ket{\Psi_{t}}\bra{\Psi_{t}})  \otimes \ket{0}_{t+1} \bra{0} \big) \mathcal{G}_{t, t+1}^\dagger) \nonumber \\
            &= \Tr_{t} (\mathcal{G}_{t, t+1} \big( \rho_{t}  \otimes \ket{0}_{t+1} \bra{0} \big) \mathcal{G}_{t, t+1}^\dagger) .
\end{align}
To keep notation simple we have written again $\mathcal{G}_{t,t+1}$ for the unitaries emerging from the original global update operator by suppressing the identities on certain columns. Similarly, we have denoted by $\ket{\Psi_{t }}$ the state of the QCA after $t$ applications of the global update although it is only supported up to column $t$. In order to arrive at this recurrence relation, we exploited that the global updates are the same, always acting on two contiguous time-columns only, and that we initialize all sites apart from those in the first column in a same (vacant) state. Note that while the update for the total system is unitary, the one on the level of the reduced states in Eq.~\eqref{eq1} is, in general, not. In the following Sections we want to investigate how the global update determines the form of the map which transforms a reduced state according to Eq.~\eqref{eq1}. We will show that through a decomposition 
\begin{equation} \label{eq2}
    \mathcal{G}_{t,t+1} = \Bigg( \prod_{k=1}^{m-1} L_k \Bigg) \Bigg( \prod_{k=m}^{N} G_k \Bigg) \Bigg(  \prod_{k=1}^{m-1} R_k \Bigg) ,
\end{equation}
in terms of local unitary operators $L_k,G_k,R_k$, the dynamics of the (1+1)D QCA can approximate any local Markovian open quantum time evolution. While the operators $L_k$ act on one \textit{control} site in column $t$ and one \textit{target} site in column $t+1$, the operators $G_k$ and $R_k$ act on $m$ control sites and one target site (cf. Fig.~\ref{qca}). Their explicit form is given in the next Section. \\
In Ref.~\cite{gillman2022b} it was shown that the theoretical framework of (1+1)D QCA is equivalent to the one of (deep) dissipative QNNs \cite{beer2020}. The organization of the QCA into columns of finite dimensional quantum systems which get successively updated from left to right naturally allows to identify the first column as an input layer, the last column as an output layer and the remaining ones as hidden layers. Indeed, the state of the first layer can be chosen freely and the result of a computation---propagating along the time axis according to Eq.~\eqref{eq1}---is represented by the reduced state of the final layer. The fundamental building blocks of our (1+1)D QCA dynamics, i.~e.~the local gates in a decomposition like in Eq.~\eqref{eq2}, correspond to the perceptrons of an (in general recurrent) QNN. Note also that output and hidden layers are initialized in an all-vacant reference state. While being structurally equivalent, (1+1)D QCA and QNNs differ in their field of application. The former deal with the controlled simulation of many-body quantum dynamics whereas the latter are quantum approaches to pattern recognition in data samples. Exploiting the link between them, our results give analytical and numerical insights into both fields.
\section{Lindbladian evolution of the initial state}
The purpose of this Section is to find the explicit forms of the local gates $L_k(\delta t), G_k(\delta t), R_k(\delta t)$ such that the global gate in Eq.~\eqref{eq2} gives rise to the most general Lindbladian evolution of the reduced layer state with local Hamiltonians and dissipation processes (cf. Fig.~\ref{qca}). More precisely, we want to show that for small $\delta t$ the gate in Eq.~\eqref{eq2} propagates the reduced state at time $t$ according to 
\begin{equation} \label{eq3}
    \rho(t+1)\approx \rho(t) + \delta t \mathcal{L} [\rho (t)] \approx e^{\delta t \mathcal{L}}[\rho(t)] 
\end{equation}
with the \textit{Lindblad generator} \cite{lindblad1976, gorini1976, breuer2002}
\begin{align} \label{lindbladgenerator}
    \mathcal{L}[\rho] &= -i [\sum_{{\beta=1}}^{4^m-1}\sum_{k=1}^N H_k^{[\beta]}, \rho] + \sum_{{\beta=1}}^{4^m-1} \sum_{k=1}^{N}  \gamma_k^{[\beta]} \bigg(A_k^{[\beta]} \rho \Big( A_k^{[\beta]} \Big) ^\dagger - \frac{1}{2} \Big\{ \Big( A_k^{[\beta]} \Big) ^\dagger A_k^{[\beta]} , \rho \Big\} \bigg) , \nonumber  \\
    &= \sum_{{\beta=1}}^{4^m-1}\sum_{k=1}^N \mathcal{L}_k^{[\beta]}[\rho]
\end{align}
the $m$-local Hamiltonians
\begin{align}\label{Hamiltonians}
    H_k^{[\beta]}=\Tilde{H}_{k-m+1,k-m+2,\ldots, k}^{[\beta]} =& \sum_{\alpha_1,\alpha_2, \ldots, \alpha_m} d^{[\beta], \alpha_1,\alpha_2, \ldots, \alpha_m}_{k} \sigma^{\alpha_1}_{k-m+1} \otimes \sigma^{\alpha_2}_{k-m+2} \otimes \cdots \otimes \sigma^{\alpha_m}_{k} \, ,
\end{align}
and the $m$-local \textit{jump operators} 
\begin{equation}\label{jump_op}
    A_{k}^{[\beta]}=\Tilde{A}_{k-m+1,k-m+2,\ldots, k}^{[\beta]}  = \sum_{\alpha_1, \alpha_2, \ldots, \alpha_m} c_{k}^{[\beta], \alpha_1, \alpha_2, \ldots, \alpha_m} \sigma_{k-m+1}^{\alpha_1} \otimes \sigma_{k-m+2}^{\alpha_2} \otimes \cdots \otimes \sigma_{k}^{\alpha_m}.
\end{equation}
Note that, in contrast to the last Section, in Eq.~\eqref{eq3} we refer to the time as an argument rather than as an index. This is because we want to distinguish in the following between the instantaneous reduced state at a certain time and the layer on which this is supported. The former will be indicated as an argument while for the latter we use an index notation. In Eqs.~\eqref{Hamiltonians}, \eqref{jump_op} the symbol $\sigma^{\alpha= 1,2,3}$ denotes the Pauli matrices and $\sigma^0$ the identity. The indices $\alpha_1,\ldots, \alpha_m$ are summed over $0,1,2,3$. We consider first the general case of periodic boundary conditions with
\begin{align*}
    &\sigma_{k-m+1}^{\alpha_1} \otimes \sigma_{k-m+2}^{\alpha_2} \otimes \cdots \otimes \sigma_{0}^{\alpha_i} \otimes \sigma_{1}^{\alpha_{i+1}} \otimes \cdots \otimes \sigma_{k}^{\alpha_m} \\
    =& \sigma_{N+k-m+1}^{\alpha_1} \otimes \sigma_{N+k-m+2}^{\alpha_2} \otimes \cdots \otimes \sigma_{N}^{\alpha_i} \otimes \sigma_{1}^{\alpha_{i+1}} \otimes \cdots \otimes \sigma_{k}^{\alpha_m} , 
\end{align*}
implying that sites at the ends may interact locally. Choosing nonnegative rates $\gamma_k^{[\beta]}$, the coefficients $c_k^{[\beta],\alpha_1,\alpha_2,\ldots,\alpha_m}$ such that the operators $A_{k}^{[\beta]}$ are pairwise orthonormal and the coefficients $d_k^{[\beta],\alpha_1,\alpha_2,\ldots,\alpha_m}$ such that the $H_k^{[\beta]}$ are Hermitian, Eq.~\eqref{lindbladgenerator} implements the most general Lindbladian dynamics with jump operators and local Hamiltonians supported on at most $m$ sites. Motivated by Eqs.~\eqref{Hamiltonians}, \eqref{jump_op} we define the operators acting on layer $t$
\begin{align*}
    H_{k,t}^{[\beta]}=& \sum_{\alpha_1,\alpha_2, \ldots, \alpha_m} d^{[\beta],\alpha_1,\alpha_2, \ldots, \alpha_m}_{k} \sigma^{\alpha_1}_{k-m+1,t} \otimes \sigma^{\alpha_2}_{k-m+2,t} \otimes \cdots \otimes \sigma^{\alpha_m}_{k,t}  \\
    A_{k,t}^{[\beta]} =& \sum_{\alpha_1, \alpha_2, \ldots, \alpha_m} c_{k}^{[\beta], \alpha_1, \alpha_2, \ldots, \alpha_m} \sigma_{k-m+1,t}^{\alpha_1} \otimes \sigma_{k-m+2,t}^{\alpha_2} \otimes \cdots \otimes \sigma_{k,t}^{\alpha_m}
\end{align*}
and furthermore the Hermitian operator acting on layers $t, t+1$
\begin{equation*}
    V_{k,(t,t+1)}^{[\beta]} = \tilde{V}_{k-m+1,k-m+2, \ldots, k, (t,t+1)}^{[\beta]} = \sqrt{\gamma_k^{[\beta]}} \Big( A_{k,t}^{[\beta]} \otimes \sigma^+_{k,t+1} + \Big( A_{k,t}^{[\beta]} \Big)^\dagger \otimes \sigma_{k, t+1}^-  \Big)
\end{equation*}
with the raising and lowering operators $\sigma^\pm = (\sigma^1 \pm i \sigma^2)/2$. We also introduce the swap gate
\begin{equation*}
    \mathrm{SWAP}_k = \frac{\sum_{\alpha=0}^3 \sigma_{k,t}^\alpha \otimes \sigma_{k,t+1}^\alpha}{2}.
\end{equation*}
Given these operators, we argue below that the choice [see Eq.~\eqref{eq2}]
\begin{align} \label{lgr}
    L_k =&  \mathrm{SWAP}_k\, ,\\
    G_k =&  \mathrm{SWAP}_k e^{-i \sqrt{\delta t} V_{k,(t,t+1)}^{[\beta]}} e^{-i \delta t H_{k,t}^{[\beta]}} \,,\\
    R_k =&   e^{-i \sqrt{\delta t} V_{k,(t,t+1)}^{[\beta]}} e^{-i \delta t H_{k,t}^{[\beta]}}\,,
\end{align}
evolves the reduced state according to Eq.~\eqref{eq3}. The gates $G_k,R_k$ all have the same structure and furthermore become translation invariant if 
\begin{align*}
    \gamma_1^{[\beta]} &= \gamma_2^{[\beta]} = \cdots = \gamma_N^{[\beta]}, \\
    d^{[\beta],\alpha_1,\alpha_2, \ldots, \alpha_m}_{1} &= d^{[\beta],\alpha_1,\alpha_2, \ldots, \alpha_m}_{2} = \cdots = d^{[\beta],\alpha_1,\alpha_2, \ldots, \alpha_m}_{N}, \\
    c^{[\beta],\alpha_1,\alpha_2, \ldots, \alpha_m}_{1} &= c^{[\beta],\alpha_1,\alpha_2, \ldots, \alpha_m}_{2} = \cdots = c^{[\beta],\alpha_1,\alpha_2, \ldots, \alpha_m}_{N} .
\end{align*}
First, we reorder the gate $\mathcal{G}_{t,t+1}$ such that all swap operators are collected to the left. This can be done since each swap gate in the $G_k$ is multiplied from the left only by operators with different support and therefore commutes with them. Then, expanding the operators $e^{-i\sqrt{\delta t} V_{k,(t,t+1)}^{[\beta]}}$ to second order in $\sqrt{\delta t}$, the operators $e^{-i \delta t H_{k,t}^{[\beta]}}$ to first order in $\delta t$ and, in their products, keeping only terms up to second order in $\sqrt{\delta t}$, the global gate reads
\begin{align*}
    \mathcal{G}_{t,t+1} =& \prod_{k=1}^N \mathrm{SWAP}_k \Bigg[ \mathds{1} -i \sqrt{\delta t} \sum_{k=1}^N V_{k,(t,t+1)}^{[\beta]} - \frac{\delta t}{2} \sum_{k=1}^N \big( V_{k,(t,t+1)}^{[\beta]} \big)^2  -i \delta t \sum_{k=1}^N H_{k,t}^{[\beta]}  \\
    & - \delta t \sum_{\substack{k,k' \in \{ m,\ldots,N \} \\ k < k'}} V_{k,(t,t+1)}^{[\beta]} V_{k',(t,t+1)}^{[\beta]} - \delta t \sum_{\substack{k,k' \in \{ 1,\ldots,m-1 \} \\ k < k'}} V_{k,(t,t+1)}^{[\beta]} V_{k',(t,t+1)}^{[\beta]} \\
    &- \delta t \sum_{k=m}^N \sum_{k'=1}^{m-1} V_{k,(t,t+1)}^{[\beta]} V_{k',(t,t+1)}^{[\beta]} + \mathcal{O}(\delta t^{3/2}) \Bigg] .
\end{align*}
Inserting in Eq.~\eqref{eq1} an identity operator, in the form $\mathds{1}=\prod_k {\rm SWAP}_k^2$, between the $\mathcal{G}_{t,t+1}$ and the state, the propagation prescription can then be written as
\begin{align*}
    \rho_{t+1}(t+1) = \Tr_t \Bigg( \mathcal{G}_{t+1, t}^{(2)} \ket{0}_t\bra{0} \otimes \rho_{t+1}(t)  \bigg( \mathcal{G}_{t+1, t}^{(2)} \bigg)^\dagger \Bigg) + \mathcal{O}(\delta t^{3/2}) ,
\end{align*}
with $\mathcal{G}_{t+1, t}^{(2)}$ the second order contribution to $\mathcal{G}_{t,t+1}$ where layers $t$ and $t+1$ are swapped. Denoting by $V_{k,(t+1,t)}^{[\beta]}$ and $H_{k,t+1}^{[\beta]}$ the operators $V_{k,(t,t+1)}^{[\beta]}$ and $H_{k,t}^{[\beta]}$ with swapped layers, respectively, and again keeping only contributions up to second order in $\sqrt{\delta t}$, we compute
\begingroup
\allowdisplaybreaks
\begin{align*}
    \rho_{t+1}(t+1) =& \rho_{t+1}(t) -i\delta t \big[ \sum_{k=1}^N H_{k,t+1}^{[\beta]}, \rho_{t+1}(t) \big] \\
    &-i \sqrt{\delta t} \Tr_t \big( \big[ \sum_{k=1}^N V_{k,(t+1,t)}^{[\beta]},\ket{0}_t\bra{0} \otimes \rho_{t+1}(t) \big] \big) \\
    &-\frac{\delta t}{2}\Tr_t \big( \big\{ \sum_{k=1}^N \big( V_{k,(t+1,t)}^{[\beta]} \big)^2,\ket{0}_t\bra{0} \otimes \rho_{t+1}(t) \big\} \big) \\
    &+ \delta t \Tr_t \big( \sum_{k=1}^N V_{k,(t+1,t)}^{[\beta]} \ket{0}_t \bra{0} \otimes \rho_{t+1}(t) \sum_{k=1}^N V_{k,(t+1,t)}^{[\beta]} \big)\\
    &-\delta t\Tr_t \big( \big\{ \sum_{\substack{k,k' \in \{ m,\ldots,N \} \\ k < k'}} V_{k,(t+1,t)}^{[\beta]} V_{k',(t+1,t)}^{[\beta]},\ket{0}_t\bra{0} \otimes \rho_{t+1}(t) \big\} \big)\\
    &-\delta t\Tr_t \big( \big\{ \sum_{\substack{k,k' \in \{ 1,\ldots,m-1 \} \\ k < k'}} V_{k,(t+1,t)}^{[\beta]} V_{k',(t+1,t)}^{[\beta]},\ket{0}_t\bra{0} \otimes \rho_{t+1}(t) \big\} \big)\\
    &-\delta t\Tr_t \big( \big\{ \sum_{k = m}^N \sum_{k'=1}^{m-1} V_{k,(t+1,t)}^{[\beta]} V_{k',(t+1,t)}^{[\beta]},\ket{0}_t\bra{0} \otimes \rho_{t+1}(t) \big\} \big) 
    + \mathcal{O}(\delta t^{3/2}) .
\end{align*}
\endgroup
While we immediately recognize the coherent contribution to the Lindblad master equation, it remains to be shown that the last terms have the form of a dissipator. Since $\Tr (\sigma_k^\pm \ket{0}\bra{0})=0 \ \forall k$ we have 
\begin{equation*}
    -i \sqrt{\delta t} \Tr_t \big( \big[ \sum_{k=1}^N V_{k,(t+1,t)}^{[\beta]},\ket{0}_t\bra{0} \otimes \rho_{t+1}(t) \big] \big) = 0.
\end{equation*}
Similarly, since $\Tr (\sigma_k^\pm \sigma_{k'}^\pm \ket{0}\bra{0})=\Tr (\sigma_k^+ \sigma_{k'}^- \ket{0}\bra{0})=0 $ and $\Tr (\sigma_k^- \sigma_{k'}^+ \ket{0}\bra{0}) = \delta^{k,k'} \ \forall k , k',$ the other terms evaluate to 
\begin{align*}
    &-\frac{\delta t}{2}\Tr_t \big( \big\{ \sum_{k=1}^N \big( V_{k,(t+1,t)}^{[\beta]} \big)^2,\ket{0}_t\bra{0} \otimes \rho_{t+1}(t) \big\} \big) = -\frac{\delta t}{2} \sum_{k=1}^N \gamma_k^{[\beta]} \big\{ \Big( A_{k,t+1}^{[\beta]} \Big)^\dagger A_{k,t+1}^{[\beta]}  , \rho_{t+1}(t) \big\} \\
    &\delta t \Tr_t \big( \sum_{k=1}^N V_{k,(t+1,t)}^{[\beta]} \ket{0}_t \bra{0} \otimes \rho_{t+1}(t) \sum_{k=1}^N V_{k,(t+1,t)}^{[\beta]} \big) =  \delta t \sum_{k=1}^N \gamma_k^{[\beta]} A_{k,t+1}^{[\beta]} \rho_{t+1}(t) \Big( A_{k,t+1}^{[\beta]} \Big)^\dagger\\
    &-\delta t\Tr_t \big( \big\{ \sum_{\substack{k,k' \in \{ m,\ldots,N \} \\ k < k'}} V_{k,(t+1,t)}^{[\beta]} V_{k',(t+1,t)}^{[\beta]},\ket{0}_t\bra{0} \otimes \rho_{t+1}(t) \big\} \big) = 0\\
    &-\delta t\Tr_t \big( \big\{ \sum_{\substack{k,k' \in \{ 1,\ldots,m-1 \} \\ k < k'}} V_{k,(t+1,t)}^{[\beta]} V_{k',(t+1,t)}^{[\beta]},\ket{0}_t\bra{0} \otimes \rho_{t+1}(t) \big\} \big) = 0\\
    &-\delta t\Tr_t \big( \big\{ \sum_{k = m}^N \sum_{k'=1}^{m-1} V_{k,(t+1,t)}^{[\beta]} V_{k',(t+1,t)}^{[\beta]},\ket{0}_t\bra{0} \otimes \rho_{t+1}(t) \big\} \big) = 0.
\end{align*}
As a consequence, we see that 
    \begin{align*}
    \rho_{t+1}(t+1) 
    =& \rho_{t+1}(t) -i\delta t \big[ \sum_{k=1}^N H_{k,t+1}^{[\beta]}, \rho_{t+1}(t) \big] \\
    &+ \delta t \sum_{k=1}^N \gamma_k^{[\beta]} \Bigg( A_{k,t+1}^{[\beta]} \rho_{t+1}(t) \Big( A_{k,t+1}^{[\beta]} \Big)^\dagger -\frac{1}{2}  \bigg\{ \Big( A_{k,t+1}^{[\beta]} \Big)^\dagger A_{k,t+1}^{[\beta]}  , \rho_{t+1}(t) \bigg\} \Bigg) \\
    &+ \mathcal{O}(\delta t^{3/2}) \\
    =& \rho_{t+1}(t) + \delta t \mathcal{L}^{[\beta]}[\rho_{t+1}(t)] + \mathcal{O}(\delta t^{3/2})
\end{align*}
with $\mathcal{L}^{[\beta]}[\rho]=\sum_{k=1}^N \mathcal{L}_k^{[\beta]}[\rho]$. Thus, up to second order in $\sqrt{\delta t}$ the instantaneous reduced state at time $t+1$, defined on layer $t+1$, is given through the instantaneous reduced state at time $t$, defined on layer $t+1$, by means of an open quantum dynamics generated by $\mathcal{L}^{[\beta]}$. Hence, $\mathcal{G}_{t,t+1}$ can be seen as first swapping $\rho_t (t)$ to layer $t+1$ and then evolving it. Keeping in mind that the reduced state actually propagates successively from left to right, with each time-step implementing a change according to a Lindbladian, the updated state can be expressed as in Eq.~\eqref{eq3}. It follows that the small parameter $\delta t$ can be interpreted as a time-increment between layers $t,t+1$ (which are in units of $\delta t$) and the state $\rho(t)$ obeys the Markovian quantum master equation
\begin{align*}
    \frac{\rho (t+1) - \rho (t)}{\delta t} \approx \mathcal{L}^{[\beta]}[\rho (t)] .
\end{align*}
The current description allows us to  consider $N$ jump operators and local Hamiltonians. However, in the most general $m$-local Lindbladian [cf.~Eqs.~\eqref{lindbladgenerator}] we have $ N \cdot (4^{m}-1)$ of them occurring simultaneously (although there will be redundancies). In order to account for all of these $\mathcal{L}^{[\beta]}$, we have to introduce more vacant states and release the constraint that the gate $\mathcal{G}_{t,t+1}$ has to be the same for all $t$. As already suggested by our labeling, we denote by $\mathcal{G}_{t,t+1}^{(\beta)}$ the $4^{m}-1$ layer-to-layer unitaries yielding the different contributions to the Lindblad generator. The dynamics is then modified by requiring that these gates are repeatedly applied in ascending order from left to right. As a result, the recurrence relation for the evolution of the reduced state is now given by ($f=4^{m}-1$)
\begin{align} \label{moreprocesses}
    & \rho_{t+f}(t+f) \nonumber \\ 
    &= \Tr_{t+f-1} (\mathcal{G}_{t+f-1, t+f}^{(f)} \big( \cdots \Tr_{t}( \mathcal{G}_{t, t+1}^{(1)} \rho_{t}(t) \otimes \ket{0}_{t+1} \bra{0} {\mathcal{G}_{t, t+1}^{(1)}}^\dagger) \cdots  \otimes \ket{0}_{t+f} \bra{0} \big) {\mathcal{G}_{t+f-1, t+f}^{(f)}}^\dagger) \nonumber \\
    &\approx \rho_{t+f} (t) + \sum_{\beta=1}^{f} \mathcal{L}^{[\beta]}[\rho_{t+f} (t)]\delta t 
\end{align}
with $\mathcal{L}[\rho] = \sum_{\beta =1}^{4^m-1} \mathcal{L}^{[\beta]}[\rho]$ the most general Lindbladian for $m$-local dissipation processes and Hamiltonian which is a sum of $m$-local Hamiltonians. In the last line of Eqs.~\eqref{moreprocesses} we used again that we may neglect all contributions of higher order in $\sqrt{\delta t}$ than two. Although the reduced state of a layer is still determined by the reduced state of the previous layer according to Eq.~\eqref{eq1}, we are now interested in it only after $4^{m}-1$ iterations as incorporating $(4^{m}-1)N$ jump operators and Hamiltonians requires $4^{m}-1$ consecutive time-steps. Therefore, through $\gamma_k^{[\beta]}, d_k^{[\beta],\alpha_1, \ldots, \alpha_m}$ and $c_k^{[\beta],\alpha_1,\ldots,\alpha_m}$ we have complete parametric control over all local Lindbladian evolutions. \\
We explained how the gates $\mathcal{G}_{t,t+1}^{[\beta]}$ [see Eqs.~\eqref{eq2},\eqref{lgr}] give rise to a Lindbladian dynamics in terms of periodic boundary conditions. For open boundary conditions the Hamiltonians $H_k^{[\beta]}$ and jump operators $A_k^{[\beta]}$ in the superoperator $\mathcal{L}$ in Eq.~\eqref{lindbladgenerator} cannot be supported across the boundary and all $d_k^{[\beta],\alpha_1,\ldots,\alpha_m}$, $c_k^{[\beta],\alpha_1,\ldots,\alpha_m}$ can be set to zero in the gates $\mathcal{G}_{t,t+1}^{[\beta]}$ if $k<m$ and it reduces to
\begin{equation*}
    \mathcal{G}_{t,t+1} = \bigg( \prod_{k=1}^{m-1} \mathrm{SWAP}_k \bigg) \bigg( \prod_{k=m}^N \mathrm{SWAP}_k e^{-i \sqrt{\delta t} V_{k,(t,t+1)}^{[\beta]}} e^{-i \delta t H_{k,t}^{[\beta]}} \bigg).
\end{equation*}
Following a similar  derivation, we can show how this then results in the most general $m$-local Lindbladian evolution with open boundary conditions.

\section{Numerical simulation results}
\begin{figure}[t] 
    \centering
    \includegraphics[width=\textwidth]{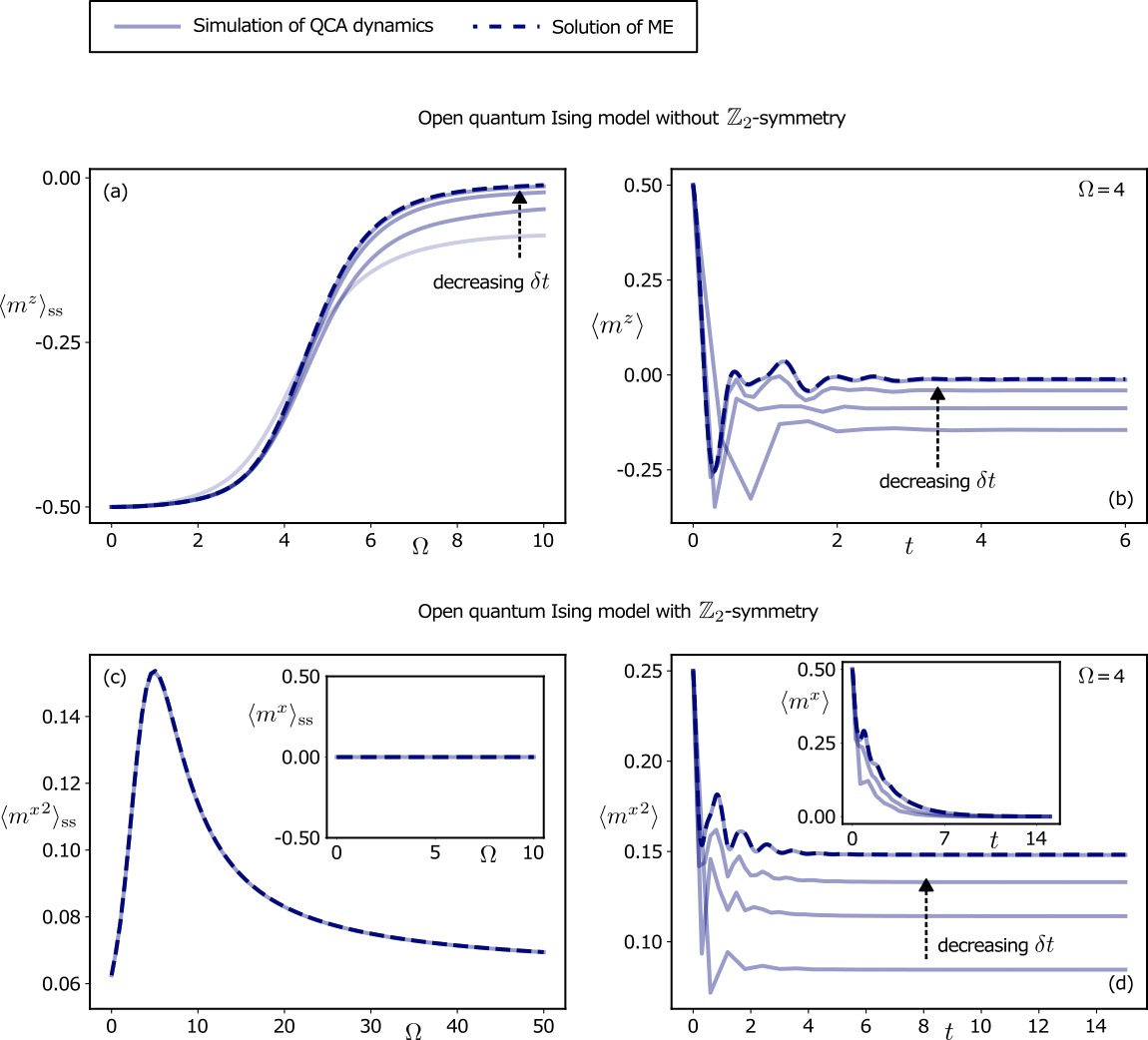}
    \caption{\textbf{Simulation of (1+1)D QCA dynamics and solution of master equation: transverse-field Ising models with local decay.} 
    (a) Stationary average magnetization in $z$-direction as function of $\Omega$ for the Ising model with Hamiltonian in Eq.~\eqref{Ising}. The curves for the numerical simulation of the QCA dynamics approach the one corresponding to the numerical solution of the master equation as $\delta t$ gets smaller. We consider the values $\delta t \approx 0.3, 0.15 , 0.043, 0.008$. The total evolution time is $t_f =30$ and we start from an all-occupied initial state. (b) For a specific field-strength $\Omega = 4$ we further compare the time-evolution of the master equation solution with the simulations of the QCA dynamics for $\delta t \approx 0.4, 0.3, 0.12, 0.008 $. (c) For the Ising model with Hamiltonian \eqref{Z2} we plot the stationary magnetization in $x$-direction (inset) and its stationary variance as a function of $\Omega$ for the solution of the master equation and the simulation of the QCA dynamics with $\delta t = 0.015$ ($\delta t = 0.08$). (d) Finally, again for $\Omega = 4$, starting from a flat initial state (see main text) we illustrate the convergence of these quantities for the QCA simulation ($\delta t \approx 0.6, 0.3, 0.2, 0.008$ and $\delta t \approx 0.6, 0.3, 0.04$ in inset), as a function of time, to the one obtained from numerically integrating the master equation. In all plots we fixed $N=4$, $V=10$ and all parameters are given in units of the dissipation rate $\kappa$ and time is given in units of $1/\kappa$.}
    \label{results1}
\end{figure}

Having established that the gate in Eq.~\eqref{eq2} can propagate the initial state of the (1+1)D QCA according to a Lindbladian as given in Eq.~\eqref{lindbladgenerator}, we now want to demonstrate that, indeed, in the limit $\delta t \to 0$ the dynamical behavior of Markovian open quantum ensembles of $N$ two-level systems can be accurately simulated (classically) by a (1+1)D QCA. To this end, we consider three concrete models whose time-evolutions are governed by Lindblad master equations (with periodic boundary conditions). They all have two-site local Hamiltonians and at most two-site jump operators, such that for each of these we can specify the parameters $d^{[\beta],\alpha_1,\alpha_2, \ldots, \alpha_m}_{k}, \gamma^{[\beta]}_k, c^{[\beta], \alpha_1,\alpha_2, \ldots, \alpha_m}_{k}$ and choose a certain total evolution time $t_{f}$ and number of layers $L$. The latter corresponds to fixing $\delta t = t_f / L$. Then we numerically simulate the respective (1+1)D QCA dynamics of the order parameter by resorting to Eq.~\eqref{eq1} with the gate 
\begin{equation*}
    \mathcal{G}_{t,t+1} = \mathrm{SWAP}_1 \Bigg( \prod_{k=2}^N \mathrm{SWAP}_k e^{-i\sqrt{\delta t} V_k^{t,t+1}} e^{-i\delta t H_{k,t}} \Bigg) e^{-i\sqrt{\delta t} V_1^{t,t+1}} e^{-i\delta t H_{1,t}}
\end{equation*}
where 
\begin{equation*}
H_{k,t} = \tilde{H}_{k-1,k,t}, \qquad
V_k^{t,t+1}=\tilde{V}_{k-1,k}^{t,t+1}\, ,
\end{equation*}
and compare the results with those obtained by numerically integrating the master equation. \\
The first model we study is an open quantum version of the transverse-field Ising model \cite{ates2012, weimer2015, rose2016} [see Fig.~\ref{results1}(a-b)] with Hamiltonian given by 
\begin{equation}\label{Ising}
    H_{\mathrm{Ising}} = \frac{\Omega}{2} \sum_{k=1}^N \sigma_k^x + \frac{V}{4} \sum_{k=1}^N \sigma_{k}^z \sigma_{k+1}^z ,
\end{equation}
rates $\gamma_k^{[\beta]} = \delta^{\beta, 1}\kappa$ and jump operators $A_k^{[\beta]} = \delta^{\beta, 1} \sigma_k^-$. Thus, in the (1+1)D QCA which simulates the open dynamics the parameters in the gate $\mathcal{G}_{t,t+1}$ are given as (see also Ref.~\cite{gillman2022b})
\begin{align*}
    d^{\alpha_1,\alpha_2}_{k} &= \frac{\Omega}{2} \delta^{\alpha_1, 0} \delta^{\alpha_2, 1} + \frac{V}{4} \delta^{\alpha_1,3} \delta^{\alpha_2,3} , \\
    \gamma_k &= \kappa , \\
    c^{ \alpha_1,\alpha_2}_{k} &= \frac{1}{2} \delta^{\alpha_1,0} \delta^{\alpha_2,1} - \frac{i}{2} \delta^{\alpha_1,0} \delta^{\alpha_2,2}.
\end{align*}
In the thermodynamic limit this model exhibits a transition/crossover from a phase where the stationary $z$-magnetization, i.~e.~the stationary expectation value of the operator 
\begin{equation*}
    m^z = \frac{\sum_{k=1}^N \sigma^z_k}{2N},
\end{equation*}
is $\langle m^z \rangle_{\mathrm{ss}} = -0.5$ (ferromagnetic) to a phase where it is $\langle m^z \rangle_{\mathrm{ss}} = 0$ (paramagnetic), upon increasing the field strength $\Omega$. Within a mean-field analysis, it is also known to possess an emergent symmetry \cite{marcuzzi2014}.
For finite $N$ this model possesses a unique stationary state \cite{schirmer2010} and in Fig.~\ref{results1}(a) it is shown for $N=4$ how the stationary $z$-magnetization changes from the ferromagnetic value to the paramagnetic one as a function of the field strength. In the figure, we show results for interaction strength $V=10$, evolution time $t_f = 30$ (which is well in the stationary regime) and with initial state the one with all-occupied sites $\rho_0 = \bigotimes_{k=1}^N \ket{\tikzcircle[black, fill=black]{2pt}}_k\bra{\tikzcircle[black, fill=black]{2pt}}$. Plotting this quantity for decreasing parameter $\kappa \delta t \approx 0.3, 0.15 , 0.043, 0.008$  we see that the stationary $z$-magnetization of the (1+1)D QCA dynamics converges---as a function of $\Omega$---to the one obtained by solving the master equation. Moreover, focusing on a specific instance $\Omega =4\kappa$, we show in Fig.~\ref{results1}(b) the time-evolution of the order parameter. Here we numerically simulated the (1+1)D QCA dynamics for $\kappa \delta t \approx 0.4, 0.3, 0.12, 0.008 $ and as can be seen, the time-behavior of the $z$-magnetization is well-captured by it for small enough values of $\kappa \delta t$. \\
For finite $N$, the different phases and the ergodicity breaking of the Ising model manifest in metastability, i.~e.~a partial relaxation into a long-lived paramagnetic or ferromagnetic state before a decay to the true stationary state, which depends on the initial state \cite{macieszczak2016, macieszczak2021, rose2016}. It is important to note that  metastability emerges in the proximity of the crossover regime and, moreover, gets more pronounced as $N$ gets larger. In the QNN picture, such a metastable behaviour can be interpreted as partial retrieval of patterns (similar to Hopfield neural networks). From a practical perspective, indeed, provided that metastability persists for sufficiently long times, this mechanism can give rise to genuine pattern retrieval over experimentally relevant time scales. \\
Secondly, we consider the modified Hamiltonian, emerging from the Hamiltonian in Eq.~\eqref{Ising} by a change of basis, i.~e.~a rotation by $\pi / 2$ about the $y$-axis,
\begin{equation} \label{Z2}
    H_{\mathbb{Z}_2} = \frac{\Omega}{2} \sum_{k=1}^N \sigma_{k}^z - \frac{V}{4} \sum_{k=1}^N \sigma_{k}^x \sigma_{k+1}^x .
\end{equation}
Now the coefficients in the local Hamiltonians are given by 
\begin{align*}
        d^{\alpha_1,\alpha_2}_{k} &= \frac{\Omega}{2} \delta^{\alpha_1, 0} \delta^{\alpha_2, 3} - \frac{V}{4} \delta^{\alpha_1,1} \delta^{\alpha_2,1} .
\end{align*}
The jump operators remain the same such that the Lindbladian possesses a strong symmetry with respect to the operator $\prod_{k=1}^N \sigma_k^z$ \cite{buca2012, zhang2020}. In the thermodynamic limit, this $\mathbb{Z}_2$-like symmetry can be spontaneously broken and besides a paramagnetic stationary solution two ferromagnetic ones may exist. Interestingly, the phase transition can be first or second-order, depending on the interaction strength \cite{overbeck2017}. For finite systems the (unique) stationary solution is the paramagnetic one as becomes evident from the inset in Fig.~\ref{results1}(c) where we depict the stationary $x$-magnetization, i.~e.~the stationary expectation value of the average operator 
\begin{equation*}
    m^x = \frac{\sum_{k=1}^N \sigma^x_k}{2N},
\end{equation*}
as a function of the field strength for $N=4$. However, as can also be seen from Fig.~\ref{results1}(c), the stationary expectation value of the operator ${m^x}^2$ has a pronounced peak, signaling a regime of field strengths where there is a high variance in the stationary paramagnetic state. As illustrated in Fig.~\ref{results1}(c) for $\kappa \delta t \approx 0.015$ (and $\kappa \delta t \approx 0.08$ in the inset), for small $\kappa \delta t$ the (1+1)D QCA simulation displays the same stationary behavior as the solution of the master equation upon varying $\Omega$. Choosing the state 
$$\rho_0 =  \bigotimes_{k=1}^N  \ket{\psi}\bra{\psi} \quad \mathrm{with} \quad \ket{\psi}= \frac{\ket{\tikzcircle[black, fill=black]{2pt}} + \ket{\tikzcircle[black, fill=white]{2pt}}}{\sqrt{2}}$$ as initial state and $t_f = 15$, in Fig.~\ref{results1}(d) we show how the time-evolution of the order parameter and the variance approach the master equation solution when simulating the QCA dynamics for $\kappa \delta t \approx 0.6, 0.3, 0.2, 0.008$. \\
While the Ising model with Hamiltonian as in Eq.~\eqref{Ising} shows metastable behavior, a transient ergodicity breaking with one long-lived paramagnetic and two ferromagnetic states is not known for the one in Eq.~\eqref{Z2} for finite $N$. This is, however, expected \cite{rose2016} and may be useful in order to study how recognition of many patterns could work in QNNs.\\
\begin{figure}[t] 
    \centering
    \includegraphics[width=\textwidth]{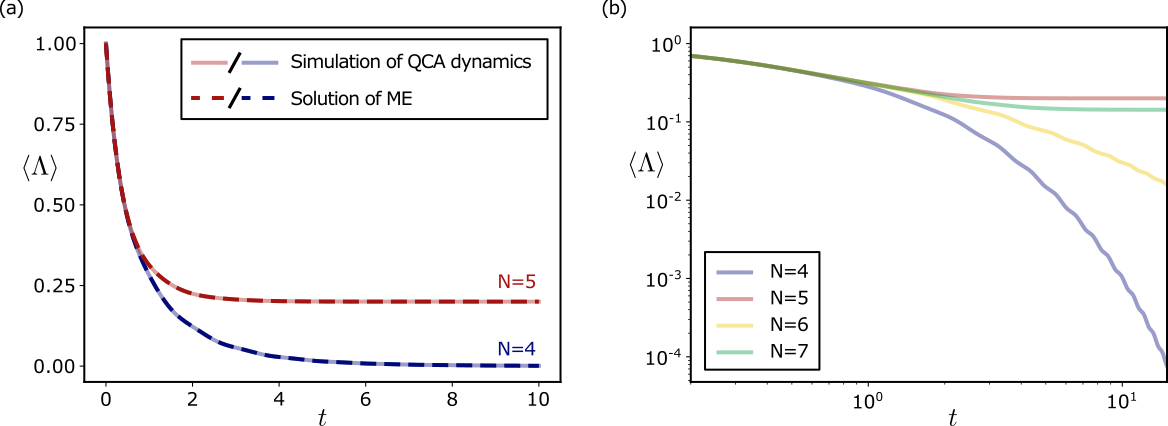}
    \caption{\textbf{Simulation of (1+1)D QCA dynamics and solution of master equation: quantum reaction-diffusion model.} (a) Starting from an all-occupied initial state, for $N=4$, $N=5$ we compare the time-behavior of the average particle density for the numerical solution of the master equation with the one obtained from simulating the QCA dynamics ($\delta t = 0.01$). (b) Simulation of the QCA dynamics for $N=4,5,6,7$ and $\delta t = 0.0015$. We have fixed $\Omega = 1$. All parameters are given in units of the dissipation rate $\kappa$ and time in units of $1/\kappa$.}
    \label{results2}
\end{figure}
Furthermore, as a model undergoing dissipation with two-site jump operators, we look at a quantum generalization of a paradigmatic reaction-diffusion process \cite{vanhorssen2015}. Namely, we study the coherent hopping Hamiltonian
\begin{equation*}
    H_{RD} = \Omega \sum_{k=1}^N \big( \sigma_k^+ \sigma_{k+1}^- + \sigma_k^- \sigma_{k+1}^+ \big)
\end{equation*}
with coupling strength $\Omega$ and (dissipative) pair-annihilation at rates $\gamma_k^{[\beta]} = \delta^{\beta, 1}\kappa$ with jump operators $A_k^{[\beta]} = \delta^{\beta, 1} \sigma_k^- \sigma_{k+1}^-$. This model features a trivial stationary state but the approach towards it is characterized by cooperative behavior \cite{perfetto2022}. Simulating its (1+1)D QCA dynamics with the parameters
\begin{align*}
    d^{\alpha_1,\alpha_2}_{k} &= \frac{\Omega}{2} ( \delta^{\alpha_1, 1} \delta^{\alpha_2, 1} +  \delta^{\alpha_1,2} \delta^{\alpha_2,2} ) \\
    \gamma_k &= \kappa \\
    c^{ \alpha_1,\alpha_2}_{k} &= \frac{1}{4} ( \delta^{\alpha_1,1} \delta^{\alpha_2,1} - i \delta^{\alpha_1,1} \delta^{\alpha_2,2} -i \delta^{\alpha_1,2} \delta^{\alpha_2,1} - \delta^{\alpha_1,2} \delta^{\alpha_2,2})
\end{align*}
we observe that the time-evolution of the expectation value of 
\begin{equation*}
    \Lambda = \frac{\sum_{k=1}^N n_i}{N}
\end{equation*}
with $n_i = \ket{\tikzcircle[black, fill=black]{2pt}}_k \bra{\tikzcircle[black, fill=black]{2pt}}$, which represents the average number of occupied states, approaches the one of the solution of the master equation. In Fig.~\ref{results2}(a), setting $\Omega = \kappa$ and starting from an all-occupied initial state, for $N=4$ and $N=5$ we compare the latter with the QCA simulation results for $\kappa \delta t = 0.01$. Already for large values of $\kappa \delta t$ we see a good agreement. In addition, simulating the QCA dynamics for $N=4,5,6,7$ with $\kappa \delta t = 0.0015$, the density of occupied states shows the characteristic splitting of the stationary states into zero and finite density ones for even and odd number of sites, respectively \cite{vanhorssen2015} [see Fig.~\ref{results2}(b)]
\section{Discussion} \label{discussion}
In this paper we have shown how a general many-body Lindbladian dynamics emerges in completely unitary (1+1)D QCA in the limit of a large number of layers. We further underpinned our analysis by numerically simulating the (1+1)D QCA dynamics of two variants of open quantum Ising models with local decay and an open quantum reaction-diffusion process with coherent particle hopping and dissipative pair-annihilation. This opens several paths for future research directions. One would expect emergent collective behavior in (1+1)D QCA or QNNs when the parameters are chosen close to a critical point of a many-body system. In the large-size and large-dimension limit both Ising models feature a phase transition and ergodicity breaking with the dynamics splitting into different basins of attraction; for finite size and dimension this may manifest in metastability. However, from the practical perspective this may be sufficient. Generally, the manifold of metastable states can feature decoherence-free subspaces and noiseless subsystems, where the dynamics is completely unitary \cite{plenio1997,zanardi1997a,zanardi1997b,lidar1998,zanardi2000,knill2000,viola2001,lidar2003,zanardi2014,zanardi2015,dubois2023}. These have in the past been concatenated with quantum error correcting codes to achieve fault-tolerant quantum computing \cite{lidar1999,lidar2001}.\\
The focus of this work was on showing the connection between (1+1)D QCA, QNNs and open system dynamics. In order to investigate larger systems, we will have to rely on advanced numerical methods, e.~g.~based on tensor networks \cite{gillman2021b, gillman2022a, gillman2022b}, to overcome the limitations imposed by the exponentially growing dimension of the Hilbert space with system size. In a second step these results could be exploited to investigate more closely how learning of a dissipative quantum many-body time-evolution could work in QNNs. Having many different basins of attraction shrinks the volume of parameters for suitable open quantum dynamics \cite{lewenstein2021, bodeker2022}. Thus, a learning routine may reliably find the relation underlying the data. In Ref.~\cite{beer2020} a (quantum) learning algorithm for general perceptrons was presented and the trainability of these architectures was demonstrated. It might further be interesting to study the impact of critical points in the many-body dynamics on the learning behavior. A key point here will be to circumvent so-called Barren plateau phenomena \cite{mcclean2018,marrero2021,patti2021}.\\

\ack
The research leading to these results has received funding from the Deutsche Forschungsgemeinschaft (DFG, German Research Foundation) under Project No. 449905436, as well as through the Research Unit FOR 5413/1, Grant No. 465199066. This project has also received funding from the European Union’s Horizon Europe research and innovation program under Grant Agreement No. 101046968 (BRISQ), and from EPSRC under Grant No. EP/V031201/1. FC~is indebted to the Baden-W\"urttemberg Stiftung for the financial support of this research project by the Eliteprogramme for Postdocs. IL is a member of the Machine Learning Cluster of Excellence, funded by the Deutsche Forschungsgemeinschaft (DFG, German Research Foundation) under Germany’s Excellence Strategy – EXC number 2064/1 – Project number 390727645.

\section*{References}
\bibliography{Reference}

\providecommand{\newblock}{}
\begin{thebibliography}{10}
\expandafter\ifx\csname url\endcsname\relax
  \def\url#1{{\tt #1}}\fi
\expandafter\ifx\csname urlprefix\endcsname\relax\def\urlprefix{URL }\fi
\providecommand{\eprint}[2][]{\url{#2}}

\bibitem{georgescu2014}
Georgescu I~M, Ashhab S and Nori F 2014 {\em Rev. Mod. Phys.\/}
  \href{http://dx.doi.org/10.1103/RevModPhys.86.153}{86, 153--185}

\bibitem{altman2021}
Altman E {\em et~al.\/} 2021 {\em PRX Quantum\/}
  \href{http://dx.doi.org/10.1103/PRXQuantum.2.017003}{2, 017003}

\bibitem{schuld2022}
Schuld M and Killoran N 2022 {\em PRX Quantum\/}
  \href{http://dx.doi.org/10.1103/PRXQuantum.3.030101}{3, 030101}

\bibitem{schuld2014}
Schuld M, Sinayskiy I and Petruccione F 2014 {\em Quantum Inf. Process.\/}
  \href{http://dx.doi.org/https://doi.org/10.1007/s11128-014-0809-8}{13,
  2567--2586}

\bibitem{schuld2021}
Schuld M and Petruccione F 2021 {\em {Machine Learning with Quantum
  Computers}\/} (Cham: Springer)

\bibitem{biamonte2017}
Biamonte J, Wittek P, Pancotti N, Rebentrost P, Wiebe N and Lloyd S 2017 {\em
  Nature\/} \href{http://dx.doi.org/https://doi.org/10.1038/nature23474}{549,
  195--202}

\bibitem{wilkinson2022}
Wilkinson S~A and Hartmann M~J 2022
  \href{http://dx.doi.org/https://doi.org/10.48550/arXiv.2208.06198}{arXiv:2208.06198}

\bibitem{poland2020}
Poland K, Beer K and Osborne T~J 2020
  \href{http://dx.doi.org/https://doi.org/10.48550/arXiv.2003.14103}{arXiv:2003.14103}

\bibitem{wolpert1997}
Wolpert D and Macready W 1997 {\em IEEE Trans. Evol. Comput.\/}
  \href{http://dx.doi.org/10.1109/4235.585893}{1, 67-82}

\bibitem{arrighi2019}
Arrighi P 2019 {\em Natural Computing\/}
  \href{http://dx.doi.org/https://doi.org/10.1007/s11047-019-09762-6}{18,
  885-899}

\bibitem{farrelly2020}
Farrelly T 2020 {\em {Quantum}\/}
  \href{http://dx.doi.org/10.22331/q-2020-11-30-368}{4, 368}

\bibitem{vonneumann1996}
von Neumann J 1996 {\em {Theory of Self-Reproducing Automata}\/} (Urbana:
  University of Illinois Press)

\bibitem{wolfram1983}
Wolfram S 1983 {\em Rev. Mod. Phys.\/}
  \href{http://dx.doi.org/10.1103/RevModPhys.55.601}{55, 601--644}

\bibitem{piroli2020}
Piroli L and Cirac J~I 2020 {\em Phys. Rev. Lett.\/}
  \href{http://dx.doi.org/10.1103/PhysRevLett.125.190402}{125, 190402}

\bibitem{duranthon2021}
Duranthon O and Di~Molfetta G 2021 {\em Phys. Rev. A\/}
  \href{http://dx.doi.org/10.1103/PhysRevA.103.032224}{103, 032224}

\bibitem{hillberry2021}
Hillberry L~E, Jones M~T, Vargas D~L, Rall P, Halpern N~Y, Bao N, Notarnicola
  S, Montangero S and Carr L~D 2021 {\em Quantum Sci. Technol.\/}
  \href{http://dx.doi.org/https://doi.org/10.1088/2058-9565/ac1c41}{6, 045017}

\bibitem{mlodinow2021}
Mlodinow L and Brun T~A 2021 {\em Phys. Rev. A\/}
  \href{http://dx.doi.org/10.1103/PhysRevA.103.052203}{103, 052203}

\bibitem{nigmatullin2021}
Nigmatullin R, Wagner E and Brennen G~K 2021 {\em Phys. Rev. Res.\/}
  \href{http://dx.doi.org/10.1103/PhysRevResearch.3.043167}{3, 043167}

\bibitem{piroli2021}
Piroli L, Turzillo A, Shukla S~K and Cirac J~I 2021 {\em J. Stat. Mech. Theory
  Exp.\/}
  \href{http://dx.doi.org/https://doi.org/10.1088/1742-5468/abd30f}{2021,
  013107}

\bibitem{ney2022}
Ney P~M, Notarnicola S, Montangero S and Morigi G 2022 {\em Phys. Rev. A\/}
  \href{http://dx.doi.org/10.1103/PhysRevA.105.012416}{105, 012416}

\bibitem{sellapillay2022}
Sellapillay K, Verga A~D and Di~Molfetta G 2022 {\em Phys. Rev. B\/}
  \href{http://dx.doi.org/10.1103/PhysRevB.106.104309}{106, 104309}

\bibitem{shirley2022}
Shirley W, Chen Y~A, Dua A, Ellison T~D, Tantivasadakarn N and Williamson D~J
  2022 {\em PRX Quantum\/}
  \href{http://dx.doi.org/10.1103/PRXQuantum.3.030326}{3, 030326}

\bibitem{labuhn2016}
Labuhn H, Barredo D, Ravets S, De~L{\'e}s{\'e}leuc S, Macr{\`\i} T, Lahaye T
  and Browaeys A 2016 {\em Nature\/}
  \href{http://dx.doi.org/https://doi.org/10.1038/nature18274}{534, 667--670}

\bibitem{bernien2017}
Bernien H {\em et~al.\/} 2017 {\em Nature\/}
  \href{http://dx.doi.org/https://doi.org/10.1038/nature24622}{551, 579--584}

\bibitem{kim2018}
Kim H, Park Y, Kim K, Sim H~S and Ahn J 2018 {\em Phys. Rev. Lett.\/}
  \href{http://dx.doi.org/10.1103/PhysRevLett.120.180502}{120, 180502}

\bibitem{browaeys2020}
Browaeys A and Lahaye T 2020 {\em Nat. Phys.\/}
  \href{http://dx.doi.org/https://doi.org/10.1038/s41567-019-0733-z}{16,
  132--142}

\bibitem{wintermantel2020}
Wintermantel T~M, Wang Y, Lochead G, Shevate S, Brennen G~K and Whitlock S 2020
  {\em Phys. Rev. Lett.\/}
  \href{http://dx.doi.org/10.1103/PhysRevLett.124.070503}{124, 070503}

\bibitem{ebadi2021}
Ebadi S {\em et~al.\/} 2021 {\em Nature\/}
  \href{http://dx.doi.org/https://doi.org/10.1038/s41586-021-03582-4}{595,
  227--232}

\bibitem{scholl2021}
Scholl P {\em et~al.\/} 2021 {\em Nature\/}
  \href{http://dx.doi.org/https://doi.org/10.1038/s41586-021-03585-1}{595,
  233--238}

\bibitem{lesanovsky2019}
Lesanovsky I, Macieszczak K and Garrahan J~P 2019 {\em Quantum Sci. Technol.\/}
  \href{http://dx.doi.org/10.1088/2058-9565/aaf831}{4, 02LT02}

\bibitem{gillman2020}
Gillman E, Carollo F and Lesanovsky I 2020 {\em Phys. Rev. Lett.\/}
  \href{http://dx.doi.org/10.1103/PhysRevLett.125.100403}{125, 100403}

\bibitem{gillman2021a}
Gillman E, Carollo F and Lesanovsky I 2021 {\em Phys. Rev. A\/}
  \href{http://dx.doi.org/10.1103/PhysRevA.103.L040201}{103, L040201}

\bibitem{gillman2021b}
Gillman E, Carollo F and Lesanovsky I 2021 {\em Phys. Rev. Lett.\/}
  \href{http://dx.doi.org/10.1103/PhysRevLett.127.230502}{127, 230502}

\bibitem{gillman2022a}
Gillman E, Carollo F and Lesanovsky I 2022 {\em Phys. Rev. E\/}
  \href{http://dx.doi.org/10.1103/PhysRevE.106.L032103}{106, L032103}

\bibitem{gillman2022b}
Gillman E, Carollo F and Lesanovsky I 2023 {\em Phys. Rev. E\/}
  \href{http://dx.doi.org/10.1103/PhysRevE.107.L022102}{107, L022102}

\bibitem{domany1984}
Domany E and Kinzel W 1984 {\em Phys. Rev. Lett.\/}
  \href{http://dx.doi.org/10.1103/PhysRevLett.53.311}{53, 311--314}

\bibitem{hopfield1982}
Hopfield J~J 1982 {\em Proc. Natl. Acad. Sci. USA\/}
  \href{http://dx.doi.org/https://doi.org/10.1073/pnas.79.8.2554}{79,
  2554--2558}

\bibitem{amit1985a}
Amit D~J, Gutfreund H and Sompolinsky H 1985 {\em Phys. Rev. A\/}
  \href{http://dx.doi.org/10.1103/PhysRevA.32.1007}{32, 1007--1018}

\bibitem{amit1985b}
Amit D~J, Gutfreund H and Sompolinsky H 1985 {\em Phys. Rev. Lett.\/}
  \href{http://dx.doi.org/10.1103/PhysRevLett.55.1530}{55, 1530--1533}

\bibitem{amit1987}
Amit D~J, Gutfreund H and Sompolinsky H 1987 {\em Ann. Phys.\/}
  \href{http://dx.doi.org/https://doi.org/10.1016/0003-4916(87)90092-3}{173,
  30--67}

\bibitem{nielsen2015}
Nielsen M~A 2015 {\em Neural Networks and Deep Learning\/} (Determination
  Press)

\bibitem{goodfellow2016}
Goodfellow I, Bengio Y and Courville A 2016 {\em Deep Learning\/} (MIT Press)
  \url{http://www.deeplearningbook.org}

\bibitem{killoran2019}
Killoran N, Bromley T~R, Arrazola J~M, Schuld M, Quesada N and Lloyd S 2019
  {\em Phys. Rev. Res.\/}
  \href{http://dx.doi.org/10.1103/PhysRevResearch.1.033063}{1, 033063}

\bibitem{mangini2021}
Mangini S, Tacchino F, Gerace D, Bajoni D and Macchiavello C 2021 {\em EPL\/}
  \href{http://dx.doi.org/10.1209/0295-5075/134/10002}{134, 10002}

\bibitem{bravo2022}
Bravo R~A, Najafi K, Gao X and Yelin S~F 2022 {\em PRX Quantum\/}
  \href{http://dx.doi.org/10.1103/PRXQuantum.3.030325}{3, 030325}

\bibitem{fiorelli2022}
Fiorelli E, Lesanovsky I and M{\"u}ller M 2022 {\em New J. Phys.\/}
  \href{http://dx.doi.org/10.1088/1367-2630/ac5490}{24, 033012}

\bibitem{rotondo2018}
Rotondo P, Marcuzzi M, Garrahan J~P, Lesanovsky I and M{\"u}ller M 2018 {\em J.
  Phys. A: Math. Theor.\/}
  \href{http://dx.doi.org/10.1088/1751-8121/aaabcb}{51, 115301}

\bibitem{beer2020}
Beer K, Bondarenko D, Farrelly T, Osborne T~J, Salzmann R, Scheiermann D and
  Wolf R 2020 {\em Nat. Commun.\/}
  \href{http://dx.doi.org/https://doi.org/10.1038/s41467-020-14454-2}{11, 1--6}

\bibitem{sharma2022}
Sharma K, Cerezo M, Cincio L and Coles P~J 2022 {\em Phys. Rev. Lett.\/}
  \href{http://dx.doi.org/10.1103/PhysRevLett.128.180505}{128, 180505}

\bibitem{lecun2015}
LeCun Y, Bengio Y and Hinton G 2015 {\em Nature\/}
  \href{http://dx.doi.org/https://doi.org/10.1038/nature14539}{521, 436--444}

\bibitem{carollo2019}
Carollo F, Gillman E, Weimer H and Lesanovsky I 2019 {\em Phys. Rev. Lett.\/}
  \href{http://dx.doi.org/10.1103/PhysRevLett.123.100604}{123, 100604}

\bibitem{lorenzo2017}
Lorenzo S, Ciccarello F and Palma G~M 2017 {\em Phys. Rev. A\/}
  \href{http://dx.doi.org/10.1103/PhysRevA.96.032107}{96, 032107}

\bibitem{ciccarello2017}
Ciccarello F 2017 {\em Quantum Meas. Quantum Metrol.\/}
  \href{http://dx.doi.org/10.1515/qmetro-2017-0007}{4, 53--63}

\bibitem{ciccarello2022}
Ciccarello F, Lorenzo S, Giovannetti V and Palma G~M 2022 {\em Phys. Rep.\/}
  \href{http://dx.doi.org/https://doi.org/10.1016/j.physrep.2022.01.001}{954,
  1--70}

\bibitem{cattaneo2021}
Cattaneo M, De~Chiara G, Maniscalco S, Zambrini R and Giorgi G~L 2021 {\em
  Phys. Rev. Lett.\/}
  \href{http://dx.doi.org/10.1103/PhysRevLett.126.130403}{126, 130403}

\bibitem{cattaneo2022}
Cattaneo M, Giorgi G~L, Zambrini R and Maniscalco S 2022 {\em Open Syst. Inf.
  Dyn.\/}
  \href{http://dx.doi.org/https://doi.org/10.1142/S1230161222500159}{29,
  2250015}

\bibitem{lindblad1976}
Lindblad G 1976 {\em Commun. Math. Phys.\/}
  \href{http://dx.doi.org/https://doi.org/10.1007/BF01608499}{48, 119--130}

\bibitem{gorini1976}
Gorini V, Kossakowski A and Sudarshan E~C~G 1976 {\em J. Math. Phys.\/}
  \href{http://dx.doi.org/https://doi.org/10.1063/1.522979}{17, 821--825}

\bibitem{breuer2002}
Breuer H~P and Petruccione F 2002 {\em The Theory of Open Quantum Systems\/}
  (New York: Oxford University Press)

\bibitem{rose2016}
Rose D~C, Macieszczak K, Lesanovsky I and Garrahan J~P 2016 {\em Phys. Rev.
  E\/} \href{http://dx.doi.org/10.1103/PhysRevE.94.052132}{94, 052132}

\bibitem{overbeck2017}
Overbeck V~R, Maghrebi M~F, Gorshkov A~V and Weimer H 2017 {\em Phys. Rev. A\/}
  \href{http://dx.doi.org/10.1103/PhysRevA.95.042133}{95, 042133}

\bibitem{vanhorssen2015}
van Horssen M and Garrahan J~P 2015 {\em Phys. Rev. E\/}
  \href{http://dx.doi.org/10.1103/PhysRevE.91.032132}{91, 032132}

\bibitem{ates2012}
Ates C, Olmos B, Garrahan J~P and Lesanovsky I 2012 {\em Phys. Rev. A\/}
  \href{http://dx.doi.org/10.1103/PhysRevA.85.043620}{85, 043620}

\bibitem{weimer2015}
Weimer H 2015 {\em Phys. Rev. Lett.\/}
  \href{http://dx.doi.org/10.1103/PhysRevLett.114.040402}{114, 040402}

\bibitem{marcuzzi2014}
Marcuzzi M, Levi E, Diehl S, Garrahan J~P and Lesanovsky I 2014 {\em Phys. Rev.
  Lett.\/} \href{http://dx.doi.org/10.1103/PhysRevLett.113.210401}{113, 210401}

\bibitem{schirmer2010}
Schirmer S~G and Wang X 2010 {\em Phys. Rev. A\/}
  \href{http://dx.doi.org/10.1103/PhysRevA.81.062306}{81, 062306}

\bibitem{macieszczak2016}
Macieszczak K, Guta M, Lesanovsky I and Garrahan J~P 2016 {\em Phys. Rev.
  Lett.\/} \href{http://dx.doi.org/10.1103/PhysRevLett.116.240404}{116, 240404}

\bibitem{macieszczak2021}
Macieszczak K, Rose D~C, Lesanovsky I and Garrahan J~P 2021 {\em Phys. Rev.
  Res.\/} \href{http://dx.doi.org/10.1103/PhysRevResearch.3.033047}{3, 033047}

\bibitem{buca2012}
Bu{\v{c}}a B and Prosen T 2012 {\em New J. Phys.\/}
  \href{http://dx.doi.org/10.1088/1367-2630/14/7/073007}{14, 073007}

\bibitem{zhang2020}
Zhang Z, Tindall J, Mur-Petit J, Jaksch D and Bu{\v{c}}a B 2020 {\em J. Phys.
  A: Math. Theor.\/} \href{http://dx.doi.org/10.1088/1751-8121/ab88e3}{53,
  215304}

\bibitem{perfetto2022}
Perfetto G, Carollo F, Garrahan J~P and Lesanovsky I 2022
  \href{http://dx.doi.org/https://doi.org/10.48550/arXiv.2209.09784}{arXiv:2209.09784}

\bibitem{plenio1997}
Plenio M~B, Vedral V and Knight P~L 1997 {\em Phys. Rev. A\/}
  \href{http://dx.doi.org/10.1103/PhysRevA.55.67}{55, 67--71}

\bibitem{zanardi1997a}
Zanardi P and Rasetti M 1997 {\em Phys. Rev. Lett.\/}
  \href{http://dx.doi.org/10.1103/PhysRevLett.79.3306}{79, 3306--3309}

\bibitem{zanardi1997b}
Zanardi P 1997 {\em Phys. Rev. A\/}
  \href{http://dx.doi.org/10.1103/PhysRevA.56.4445}{56, 4445--4451}

\bibitem{lidar1998}
Lidar D~A, Chuang I~L and Whaley K~B 1998 {\em Phys. Rev. Lett.\/}
  \href{http://dx.doi.org/10.1103/PhysRevLett.81.2594}{81, 2594--2597}

\bibitem{zanardi2000}
Zanardi P 2000 {\em Phys. Rev. A\/}
  \href{http://dx.doi.org/10.1103/PhysRevA.63.012301}{63, 012301}

\bibitem{knill2000}
Knill E, Laflamme R and Viola L 2000 {\em Phys. Rev. Lett.\/}
  \href{http://dx.doi.org/10.1103/PhysRevLett.84.2525}{84, 2525--2528}

\bibitem{viola2001}
Viola L, Fortunato E~M, Pravia M~A, Knill E, Laflamme R and Cory D~G 2001 {\em
  Science\/} \href{http://dx.doi.org/10.1126/science.1064460}{293, 2059--2063}

\bibitem{lidar2003}
Lidar D~A and Birgitta~Whaley K 2003 {\em Decoherence-Free Subspaces and
  Subsystems\/} (Berlin, Heidelberg: Springer Berlin Heidelberg)

\bibitem{zanardi2014}
Zanardi P and Campos~Venuti L 2014 {\em Phys. Rev. Lett.\/}
  \href{http://dx.doi.org/10.1103/PhysRevLett.113.240406}{113, 240406}

\bibitem{zanardi2015}
Zanardi P and Campos~Venuti L 2015 {\em Phys. Rev. A\/}
  \href{http://dx.doi.org/10.1103/PhysRevA.91.052324}{91, 052324}

\bibitem{dubois2023}
Dubois J, Saalmann U and Rost J~M 2023 {\em Phys. Rev. Res.\/}
  \href{http://dx.doi.org/10.1103/PhysRevResearch.5.L012003}{5, L012003}

\bibitem{lidar1999}
Lidar D~A, Bacon D and Whaley K~B 1999 {\em Phys. Rev. Lett.\/}
  \href{http://dx.doi.org/10.1103/PhysRevLett.82.4556}{82, 4556--4559}

\bibitem{lidar2001}
Lidar D~A, Bacon D, Kempe J and Whaley K~B 2001 {\em Phys. Rev. A\/}
  \href{http://dx.doi.org/10.1103/PhysRevA.63.022307}{63, 022307}

\bibitem{lewenstein2021}
Lewenstein M, Gratsea A, Riera-Campeny A, Aloy A, Kasper V and Sanpera A 2021
  {\em Quantum Sci. Technol.\/}
  \href{http://dx.doi.org/10.1088/2058-9565/ac070f}{6, 045002}

\bibitem{bodeker2022}
B{\"o}deker L, Fiorelli E and M{\"u}ller M 2022
  \href{http://dx.doi.org/https://doi.org/10.48550/arXiv.2210.07894}{arXiv:2210.07894}

\bibitem{mcclean2018}
McClean J~R, Boixo S, Smelyanskiy V~N, Babbush R and Neven H 2018 {\em Nat.
  Commun.\/}
  \href{http://dx.doi.org/https://doi.org/10.1038/s41467-018-07090-4}{9, 1--6}

\bibitem{marrero2021}
Ortiz~Marrero C, Kieferov\'a M and Wiebe N 2021 {\em PRX Quantum\/}
  \href{http://dx.doi.org/10.1103/PRXQuantum.2.040316}{2, 040316}

\bibitem{patti2021}
Patti T~L, Najafi K, Gao X and Yelin S~F 2021 {\em Phys. Rev. Res.\/}
  \href{http://dx.doi.org/10.1103/PhysRevResearch.3.033090}{3, 033090}

\end{thebibliography}

\end{document}